\documentstyle{l-aa}
\def\lsim{\mathrel{\rlap{\lower4pt\hbox{\hskip1pt$\sim$}}
\raise1pt\hbox{$<$}}}                % less than or approx. symbol
\def\gsim{\mathrel{\rlap{\lower4pt\hbox{\hskip1pt$\sim$}}
\raise1pt\hbox{$>$}}}                % greater than or approx. symbol
\newcommand{\be}{\begin{equation}}
\newcommand{\ee}{\end{equation}}
\newcommand{\Tau}{\mbox{{\Large $\tau$}}}
\newcommand{\gga}{\alpha}
\newcommand{\ggb}{\beta}
\newcommand{\ggg}{\gamma}
\newcommand{\ggm}{\mu}
\newcommand{\gge}{\epsilon}

\newcommand{\ggs}{\sigma}
\newcommand{\ggn}{\nu}
\newcommand{\ggl}{\lambda}
\newcommand{\ggt}{\tau}
\newcommand{\Gg}{\Gamma}
\newcommand{\EM}{\mbox{{\scriptsize $EM$}}}
\newcommand{\M}{\mbox{{\scriptsize $M$}}}
\newcommand{\R}{\mbox{{\scriptsize $R$}}}
\newcommand{\pder}{\partial}
\newcommand{\dpar}[2]{\frac{\pder #1}{\pder #2}}
\newcommand{\cris}[3]{\Gg^{#1}_{#2 #3}}
\newcommand{\aap}{A\&A}
\newcommand{\apj}{ApJ}
\newcommand{\apjl}{ApJ Lett.}
\newcommand{\mnras}{MNRAS}
\newcommand{\qjras}{QJRAS}
\newcommand{\prd}{Phys. Rev. D}
\begin{document}
\thesaurus{2(02.13.2; 02.18.8; 12.12.1)}
\title{Magnetic fields and large scale structure in a hot
universe. I. General equations}

\author{E. Battaner 
\and E. Florido 
\and J. Jim\'enez-Vicente}

\institute{Departamento de F\'{\i}sica Te\'orica y del Cosmos \\
	Universidad de Granada. Spain }
\maketitle
\begin{abstract}
	We consider that no mean magnetic field exists during this epoch, but
that there is a mean magnetic energy associated with large-scale magnetic
inhomogeneities. We study the evolution of these inhomogeneities and their
influence on the large scale density structure, by introducing linear 
perturbations in Maxwell equations, the conservation of momentum-energy equation,
and in Einstein field equations. The primordial magnetic field structure is
time independent in the linear approximation, only being diluted by the general expansion, so that $\vec{B}
R^2$ is conserved in comoving coordinates. 
Magnetic fields have a strong influence on the formation of large-scale structure.
Firstly, relatively low fields are able
to generate density structures even if they were inexistent at earlier times. Second,
magnetic fields act anisotropically more recently, modifying the evolution of
individual density clouds. Magnetic flux tubes have a tendency to concentrate
photons in filamentary patterns.
\end{abstract}

\section{Introduction}
%\label{sec1}
	Recent measurements of intergalactic magnetic fields (see the review
by Kronberg, 1994, and references therein) have provided evidence of the
following facts: 
\begin{enumerate}
\item {\em Magnetic fields of the order of $3\mu G$ are not uncommon.} Values
larger than $1 \mu G$ have been found in all reported measurements (see also
Feretti et al. 1995). Fields of this strength are quantitatively important
because they correspond to an energy density equal to that of CMBR.
Kronberg (1994)
even suggests that $3 \mu G$ fields are ubiquitous. In this case, a ``background
magnetism'' would be in equipartition of energy with the CMBR. This possibility,
even whilst theoretically attractive, is still based on a limited number of measurements
and therefore will not be assumed here.
\item {\em Large magnetic fields have also been found in protogalactic clouds }(e.g.
 Wolfe, Lanzetta, \& Oren 1992; Welter, Perry, \& Kronberg, 1984). Probably, 
primordial magnetic fields to a large degree contributed to the present intergalactic ones.
Kronberg suggested that magnetic fields play a role in the formation of
structures in the Universe. 
\end{enumerate}
	
These facts, even if we do not know exactly how ubiquitous and how persistent
in time intergalactic magnetic fields are, stimulate the analysis of their
evolution and interrelation with density inhomogeneities.

In this paper we consider a universe dominated by relativistic particles. More
precisely we have in mind a universe dominated by photons before Equality.
The equations however are valid for any kind of dominant relativistic particles
including hot dark matter.

	The study of the evolution of density inhomogeneities for a universe
with photons and barions, when no magnetic
fields are present, is a classical topic (Weinberg 1972;
Peebles 1980;
 Kolb \& Turner 1990; B\"{o}rner 1988;
Battaner 1996). It may be divided into three periods:
\begin{quote}
i) {\em Post-Recombination era}. During this era a Newtonian analysis is
appropriate, but nonlinear effects require rather sophisticated numerical
techniques. Inhomogeneities grow as $R$ first, becoming proportional to $R^2$
and $R^3$ when nonlinear effects become more and more important. The inclusion
of magnetic fields in the study of this epoch has been carried out by
Wasserman (1978)
, Coles (1992) and Kim, Olinto, \& Rosner (1994), this last work including
nonlinear effects. Classical treatments of Birkeland currents in the plasma
Universe have been carried out by Peratt (1988) 
\end{quote}
\begin{quote}
ii) {\em Acoustic era}. A relativistic treatment is necessary; viscosity and heat
conduction must be included (Field 1969;
Weinberg 1972; and others) as these
effects explain the Silk mass. Inhomogeneities do not increase during this era, which
ends at Recombination, its beginning being dependent on the rest mass of the
primordial cloud, around $R=10^{-5}R_{0}$. As far as we know, no attempt at
introducing magnetic fields into this analysis has been made.
\end{quote}
\begin{quote}
iii) {\em Radiation dominated era}. This era ends when the acoustic one begins,
and is therefore not perfectly defined, roughly at $R\approx 10^{-5}R_{0}$, so
it corresponds to a photon dominated universe. The beginning of this era is also
rather indeterminate. During this epoch non-magnetic inhomogeneities increase as $R^2$.
\end{quote}
The inclusion of magnetic fields in the study of this third era is the objective
of this paper.
Basically, our objective in this paper is to extend the work by Wasserman
(1978) and Kim, Olinto and Rosner (1994) to the radiation dominated era. 
We will deal with the evolution of magnetic fields and their influence
on density inhomogeneities in a radiation dominated Universe. The mathematical
procedure must be relativistic, but the inclusion of nonlinear and imperfect
fluid effects is not necessary, which greatly simplifies the problem. The
upper time boundary will be placed at approximately $10^{-5}R_{0}$, before the
Acoustic epoch, and near Equality. The lower time boundary is undefined, but 
in particular we consider a Post-Annihilation era, in order to avoid sudden
jumps in the temperature of photons and because positron-electron and quark-gluon
plasmas, which constitute the plasma state at earlier epochs, require another
analytical formulation.
Therefore, the period under study broadly extends from Annihilation to Equality.

	We consider that the evolution of magnetic fields is not perturbed by
creation and loss processes. Some mechanisms have been invoked for later
stages ( Rees 1987; Lesch \& Chiba 1995;
Ruzmaikin, Sokoloff, \& Shukurov 1989; and others)
but these probably do not affect the epoch studied here. Some mechanisms producing
primordial fields, prior to Annihilation, are implicitly assumed (see
Turner
\& Widrow 1988; Quashnock, Loeb, \& Spergel 1989;
Vachaspati 1991;
Ratra 1992;
 Enqvist \& Olesen 1993, 1994; 
Davis \& Dimopoulos 1996) but no assumption
about their order of magnitude is here adopted.

	Some important works have recently dealt with MHD in an expanding universe
(Holcomb 1989, 1990;
Dettmann, Frankel, \& Kowalenko 1993; 
Gailis et al. 1994; 
 Gailis, Frankel, \& Dettmann 1995;  
Brandenburg, Enqvist, \& Olesen 1996). However
our objective is not MHD, but the influence of magnetic fields on the formation
of large scale structure. In these papers the metric is unperturbed. Here
magnetic fields themselves are responsible for perturbations in the metric, which
induce motions and density inhomogeneities, which in turn affect the perturbed
metric, and possibly the magnetic fields.

	We have not included either protons or electrons in the system of
equations. In a first attempt to solve
this problem, this omission can be accepted, especially when we are considering
an epoch of the Universe that is dominated by photons in which charges are 
considered
to play a minor role. Nevertheless, $\nabla \times \vec{B}$ exists, which
creates a charge current (see eq. (\ref{ec})) and has an influence on the remaining 
equations. The existence of charges is implicitly assumed to support magnetic
fields and electric currents, but equations of this third component are not
explicitly considered. The influence of magnetic fields on the photon
inhomogeneities lies in the fact that curvature is decided by the
energy-momentum tensor, which is modified taking into account the magnetic
contribution. This contribution is probably too small to affect the expansion
itself, but not so as to have a large influence on the internal structure. The
contribution of magnetic fields to the energy-momentum tensor is very
different with respect to the contribution of photon and baryon densities alone.
This inclusion is not, therefore, trivial and indeed the results are clearly
different. In the presence of magnetic fields, inhomogeneities evolve in a
completely different way, as is demonstrated here in this particular epoch
in the lifetime of the Universe.

	In agreement with the Cosmological Principle, we consider than no mean
magnetic field exists at cosmological scales, so $<\vec{B}>=0$ (or rather, we
demonstrate in Appendix (\ref{app1}) that $<\vec{B}>=0$ in a Robertson-Walker 
metric). Classically, this condition is equivalently reached from $\nabla \cdot
\vec{B} = 0$, Gauss theorem and the Cosmological Principle.
However random magnetic fields do exist at lower scales in characteristic cells. These
fluctuative magnetic fields, even with random orientations, are present everywhere,
so that $<B^2>$ is non-vanishing. There is no mean magnetic field in the Universe, 
but there is a mean magnetic energy. No assumption about its value, such
as that of equipartition with the CMBR energy density or any other hypothesis, 
is made ``a priori''.

	In this paper we obtain the
equations and derive the basic conclusions. In forthcoming papers we will deal 
with inhomogeneities affected by selected particular magnetic configurations,
and with the influence of a large scale magnetic field distribution on the large scale
density distribution. Paper II deals with the influence of magnetic flux tubes
on the distribution of the density, showing that primordial magnetic fields can be
responsible for the observed present filamentary structure

The curvature has been set equal to zero, which is, in any case, a good assumption
for this epoch. 
\section{The mean magnetic field}
%\label{sec2}

	Before considering the interrelation between fluctuative magnetic fields
and density inhomogeneities, let us consider the mean magnetic field and its
influence on the motion of the Universe as a whole. It is intuitive that an
isotropic universe cannot possess a mean magnetic field. Nevertheless, it is
demonstrated in Appendix (\ref{app1}) that the existence of a mean magnetic field is
incompatible with the Robertson-Walker metric, by examining the form adopted by
Maxwell equations, the equation of motion and the Einstein field equations in a
magnetized universe. Therefore, we consider $<\vec{B}>=0$. However, magnetic fields
may be present (and actually they are) in smaller cells, with the direction of
the field being random at larger scales, so that we assume $<B^2>\neq 0$. 
There is no mean magnetic field but there is a mean magnetic energy. 

	In Einstein field equations we must include the mean magnetic
	energy-momentum tensor,
deduced from: 
\be
%\label{met}
4 \pi \Tau^{\gga \ggb}_{\EM}={\cal F}^{\gga}_{~\ggg} {\cal F}^{\ggb \ggg}-\frac{1}{4} g^{\gga 
\ggb} {\cal F}_{\ggg \ggm} {\cal F}^{\ggg \ggm}
\ee
where ${\cal F}^{\ggm \ggn}$ is the Faraday tensor. We adopt for the 
contravariant-covariant form of the Faraday tensor in a Minkowskian frame
\be
%\label{ft}
\hat{{\cal F}}^{\gga}_{~\ggg}=
\left(
\begin{array}{cccc}
0   & E_1   & E_2   & E_3   \\
E_1 &  0    & B_3   & -B_2  \\
E_2 & -B_3  &  0    & B_1   \\
E_3 & B_2   & -B_1  &  0  
\end{array}
\right)
\ee
or in brief
\be
%\label{fts}
\hat{{\cal F}}^{\gga}_{~\ggg}=
\left(
\begin{array}{cc}
0	&	\vec{E} \\
\vec{E}	&	{\cal B}
\end{array}
\right)
\ee
where
\be 
%\label{bt}
{\cal B}=
\left(
\begin{array}{ccc}
0	&	B_3	&	-B_2	\\
-B_3	&	0	&	B_1	\\
B_2	&	-B_1	&	0
\end{array}
\right)
\ee
and $\vec{E}$ and $\vec{B}$ are the electric and magnetic three-vectors, as
seen by an inertial observer. 

The Robertson-Walker metric for $k=0$ can be written as
\begin{eqnarray}
%\label{rwm}
g_{00}	&	=	&	-1 \nonumber	\\
g_{0i}	&	=	&	0  		\nonumber \\
g_{ij}	&	=	&	R^2\delta_{ij} 
\end{eqnarray}
where (as usual) Latin indexes denote only spatial coordinates, and $R$ is
the cosmological scale factor. Throughout the paper we will consider that $R$ is
measured taking its present value $R_0$ as unity. Therefore, $R$ is dimensionless,
$R=1$ at present, and we have approximately $R=z^{-1}$, with $z$ being the redshift, 
taking into account that we are dealing with times long before Recombination.
Now we take into account the transformation of coordinates that transforms
$\hat{g}_{\ggm \ggn}=diag(-1, 1, 1, 1)$ into $g_{\ggm \ggn}=diag(-1, R^2, R^2,
R^2)$. This transformation is achieved by
\be
\Lambda^{\ggm}_{~ \ggn}=\dpar{x^{\ggm}}{x'^{\ggn}}=diag(1, \frac{1}{R},
\frac{1}{R}, \frac{1}{R})
\ee
We then obtain
\be
{\cal F}^{\gga}_{~ \ggb}=\Lambda^{\gga}_{~ \ggm} \Lambda^{~ \ggn}_{\ggb} 
\hat{{\cal F}}^{\ggm}_{~ \ggn}= \left(
\begin{array}{cc}
0	& \vec{E}R \\
\frac{\vec{E}}{R} & {\cal B}
\end{array}
\right)
\ee
Using the Robertson-Walker metric we obtain the other forms of the Faraday
tensor
\begin{eqnarray}
{\cal F}_{\gga \ggb}	&	=	&	\left(
	\begin{array}{cc}
	0		&	-R\vec{E}	\\	
	R\vec{E}	&	R^2{\cal B}
	\end{array}
	\right)	\\
{\cal F}_{\gga}^{~\ggb}	&	=	&	\left(
	\begin{array}{cc}
	0		&	-\frac{\vec{E}}{R}	\\
	-R\vec{E}	&	{\cal B}
	\end{array}
	\right)	\\
{\cal F}^{\gga \ggb}	&	=	&	\left(
	\begin{array}{cc}
	0		&	R^{-1}\vec{E}	\\
	-R^{-1}\vec{E}	&	R^{-2}{\cal B}
	\end{array}
	\right)
\end{eqnarray} 

	We then assume infinite conductivity, so that electrical fields in the
rest frame of the charged particles vanish. Magnetic fields are assumed to be
tied to charges. Cheng \& Olinto (1994) showed that the effects of finite
conductivity in the early universe may be neglected.
	The electromagnetic momentum-energy tensor becomes
\be
\label{emmet}
4\pi \Tau^{\gga \ggb}_{\EM}=
\left(
\begin{array}{cccc}
0				& &	0	\\
 & & & \\
0	& &       -\frac{\vec{B}\vec{B}}{R^2}& \\
& & &
\end{array}
\right) + 
\left(
\begin{array}{cccc}
S	&	0	&	0	&	0 	\\
0	&	q	&	0	&	0	\\
0	&	0	&	q	&	0	\\
0	&	0	&	0	&	q
\end{array}
\right)
\ee
where
\be
%\label{sdef}
S=\frac{B^2}{2}
\ee
and
\be
%\label{qdef}
q=\frac{B^2}{2R^2}
\ee
We also have
\be
%\label{bpeq0}
<\vec{B}>=0
\ee
\be
%\label{bsqpdef}
<B^2_1>=\frac{<B^2>}{3}=<B^2_2>=<B^2_3>
\ee
The subindex $M$ denotes ``magnetic''
\be
\label{teim}
\Tau^{\gga \ggb}_{\M}=\left(
\begin{array}{cccc}
\frac{B^2}{8\pi}	&	0		&	0		&	0	\\
0			&	r_{\M} R^{-2}&	0		&	0	\\
0			&	0		&	r_{\M} R^{-2}&	0	\\
0			&	0		&	0		&     r_{\M} R^{-2}
\end{array}
\right)
\ee
with
\be
\label{pmdef}
r_{\M}=\frac{<B^2>}{24\pi}
\ee
This is easily checked. For instance $4\pi \Tau^{11}_{\M}=R^{-2}<-B_1
B_1+\frac{B_1 B_1}{2}+\frac{B_2 B_2}{2}+\frac{B_3 B_3}{2}>=\frac{R^{-2}}{2}
<-B_1^2+B_2^2+B_3^2>=\frac{R^{-2}}{2}<B_3^2>=\frac{R^{-2}}{6}<B^2>$. Hence 
$\Tau^{11}_{\M}=(<B^2>/24 \pi)R^{-2}$.
As
\be
\label{emdef}
\gge_{\M}=\frac{<B^2>}{8\pi}
\ee
is the magnetic energy density, we have as the equation of state
\be
\label{peos}
\gge_{\M}=3r_{\M}
\ee
	The form of the magnetic energy-momentum tensor is identical to the form
of the radiative energy-momentum tensor. For photons we consider
\be
\label{temr}
\Tau^{\gga \ggb}_{\R}=p_{\R}g^{\gga \ggb}+(\gge_{\R}+p_{\R})U^{\gga}U^{\ggb}
\ee
where $p_{\R}$ is the radiative hydrostatic pressure, $\gge_{\R}$ the radiative
energy density, and we consider $U^0=1$ and $U^i=0$. Therefore
\be
\label{temre}
\Tau^{\gga \ggb}_{\R}=\left(
\begin{array}{cccc}
\gge_{\R}	&	0	&	0	&	0	\\
0		&p_{\R}R^{-2}	&	0	&	0	\\
0		&	0	&p_{\R}R^{-2}	&	0	\\
0		&	0	&	0	&p_{\R}R^{-2}	
\end{array}
\right)
\ee
Therefore the total energy momentum is
\be
\label{temtot}
\Tau^{\gga \ggb}=\left(
\begin{array}{cccc}
\gge       &       0       &       0       &       0       \\
0               &pR^{-2}   &       0       &       0       \\
0               &       0       &pR^{-2}   &       0       \\
0               &       0       &       0       &pR^{-2}
\end{array}
\right)
\ee
where
\be
\label{etot}
\gge=\gge_{\M}+\gge_{\R}
\ee
and
\be
\label{ptot}
p=r_{\M}+p_{\R}
\ee
are the total density and hydrostatic pressure. Given the similarity of form
between $\Tau^{\gga \ggb}_{\R}$ and $\Tau^{\gga \ggb}$ the Einstein field
equations must provide familiar results, i.e. the same expansion and cooling
laws that hold for a purely radiative universe.

	It is straightforwardly obtained for the Robertson-Walker metric that
\be
\label{cs1}
\Gamma^{0}_{00}=\Gamma^0_{i0}=\Gamma^i_{00}=\Gamma^0_{ij}=\Gamma^i_{0j}=\Gamma^i_{jk}=0
\ee
\be
\label{cs2}
\Gamma^0_{ii}=R\dot{R}
\ee
\be
\label{cs3}
\Gamma^i_{0i}=\frac{\dot{R}}{R}
\ee
with $\dot{R}=\frac{dR}{dt}$, as usual, and therefore
\be
\label{epr4}
\gge \propto R^{-4}
\ee
and
\be
\label{r2per4ht}
R^2=2 \left( \frac{8\pi}{3} [ \gge R^4 ] \right)^{\frac{1}{2}} t
\ee
If $\gge_{\M} \ll \gge_{\R}$ we find the familiar result $T\propto R^{-1}$,
but in general, we must specify a relation between $\gge_{\M}$ and $\gge_{\R}$
to deduce both. For instance, if there was at some epoch equipartition of radiative
and magnetic energy $\gge_{\M} = \gge_{\R}$, both would decrease at the same
rate and the condition of equipartition would be conserved. If this condition were confirmed
at present (as suggested by Kronberg 1994) the epoch of the
Universe dominated by photons would become co-dominated equally by photons
and by magnetic fields. However, we emphasize that this equipartition condition
is not assumed (or deduced) in this paper. If it held, the relation $R \propto t^{\frac{1}{2}}$
would be maintained, but now the constant of proportionality would be higher, by a factor 
of less than $\sqrt{2}$, and the expansion would be faster.

	If we assume that the magnetic energy density is much less than the radiative
one, as is usually done, and indeed this is what our results suggest, then we would
obtain the obvious result that the expansion is unaffected by magnetic fields.

\section{The perturbed quantities}
\label{sec3}
	As usual, any quantity is substituted by its mean value plus a fluctuating
quantity, which in the linear approximation is negligible, i.e. terms including
products of two fluctuating quantities are considered second order terms. For 
instance $\gge_{\R} \rightarrow \gge_{\R}+\delta \gge_{\R}$ and $p_{\R} \rightarrow
p_{\R}+\delta p_{\R}$. As there is now no possibility of confusion we will eliminate 
the subindex $R$ and write $\gge+\delta \gge$ and $p+\delta p$.

	By considering the transformation $g_{\gga \ggb} \rightarrow g_{\gga \ggb}+
\delta  g_{\gga \ggb}$ we will call $\delta g_{\gga \ggb}=h_{\gga \ggb}$. But 
some transformations of the metric tensor do not mean real physical changes. As
argued by Weinberg (1972), it is possible to choose $h_{0i}=h_{00}=0$. We
benefit here from this choice. It is necessary to calculate $h^{\gga \ggb}_{\ast}$ 
defined as $\delta g^{\gga \ggb}$. 
The metric tensor must match $(g_{\gga \ggb} + h_{\gga \ggb})(g^{\ggb \ggg}+
h^{\ggb \ggg}_{\ast})=\delta^{\ggg}_{\gga}$ hence $\delta^{\ggg}_{\gga}+h_{\gga \ggb}
g^{\ggb \ggg}+g_{\gga \ggb}h^{\ggb \ggg}_{\ast}=\delta^{\ggg}_{\gga}$ neglecting higher
order terms. Therefore $h^{\ggb \ggg}_{\ast}=-g^{\ggb \gga} h_{\gga \ggs} g^{\ggs \ggg}$,
and we have for each component $h^{\ggb \ggg}_{\ast}=-R^{-4}h_{\ggb \ggg}$ again with
$h^{00}_{\ast}=h^{0i}_{\ast}=0$.

	$h_{\gga \ggb}$ is equivalent to a three-dimension tensor, as any
component containing the time subindex $0$ vanishes. When using three-dimension
formulae we term $h_{ij} \equiv \vec{\vec{h}}$. Its trace $h_{ii}$ plays
an important role and will simply be called $h$. We also term $h_{\ast}=
h^{ii}_{\ast}=-R^{-4}h_{ii}=-R^{-4}h$.

	$U^i$ is the four-velocity of the photon fluid. We have $\delta U^0 =0$,
as $U^0 =1$ with no perturbation. When dealing with three-dimension formulae
we term $\delta U^i \equiv \vec{u}$. As we are using comoving coordinates,
the unperturbed velocity is $U^i =0$.

	It is easily calculated that the components of the perturbed affine connection 
$\delta \Gg^{\gga}_{\ggb \ggg}$ vanish except
\begin{eqnarray}
\label{nvpac}
\delta \Gg^{i}_{jk}&=&\frac{1}{2R^2}\left[ \frac{\pder h_{ij}}{\pder x^k}+
\frac{\pder h_{ik}}{\pder x^j}-\frac{\pder h_{jk}}{\pder x^i} \right] \nonumber \\
\delta \Gg^{0}_{jk}&=&\frac{1}{2} \frac{\pder h_{jk}}{\pder t} \nonumber \\
\delta \Gg^{i}_{0j}&=&\frac{1}{2R^2}\left[ \frac{\pder h_{ij}}{\pder t}-
\frac{2 \dot{R}}{R}h_{ij} \right] 
\end{eqnarray}
	For quantities of electromagnetic nature, we would have for instance
$B_i \rightarrow B_i+\delta B_i$. But as shown in the preceding paragraph the
mean quantity is null and therefore we may use $B_i$ instead of $\delta B_i$.
Therefore $\vec{B}$, $\vec{E}$, $J^{\gga}_{\EM}$ and $\Tau^{\gga \ggb}_{\EM}$
are perturbed quantities.

\section{The perturbed Maxwell equations}
\label{sec4}
	We will obtain the perturbed equations from Maxwell's, motion, energy and 
Einstein's field equations. Not all these equations are independent but they
are all useful. The general procedure will be as usual. After substituting
any generic quantity $A$ by $A+\delta A$, the original equation is subtracted
and second order terms neglected, because we only consider linear perturbations.
As an exception, when perturbing Maxwell's equations we will not neglect second
order terms, assuming that variations in the metric and in the Faraday tensor
are uncorrelated. 
The reason is that, because the result is the same, and because some
conclusions are very important in forthcoming paragraphs, we have preferred to be
extremely cautious.

	Let us begin with the second set of Maxwell equations:
\be
\label{ssme}
{\cal F}_{\ggb \ggg ; \gga}+{\cal F}_{\ggg \gga ; \ggb}+{\cal F}_{\gga \ggb ; \ggg}=0
\ee
this yields the following perturbed equation
\begin{eqnarray}
\label{mpe}
\frac{\pder {\cal F}_{\ggb \ggg}}{\pder x^{\gga}}+
\frac{\pder {\cal F}_{\ggg \gga}}{\pder x^{\ggb}}+
\frac{\pder {\cal F}_{\gga \ggb}}{\pder x^{\ggg}}- \nonumber \\ 
\Gg^{\ggs}_{\gga \ggb}{\cal F}_{\ggs \ggg}-
\Gg^{\ggs}_{\gga \ggg}{\cal F}_{\ggb \ggs}-
\Gg^{\ggs}_{\ggb \ggg}{\cal F}_{\ggs \gga}- \nonumber \\
\Gg^{\ggs}_{\gga \ggb}{\cal F}_{\ggg \ggs}-  
\Gg^{\ggs}_{\gga \ggg}{\cal F}_{\ggs \ggb}-  
\Gg^{\ggs}_{\ggb \ggg}{\cal F}_{\gga \ggs}- \nonumber \\ 
\delta \Gg^{\ggs}_{\gga \ggb}{\cal F}_{\ggs \ggg}-
\delta \Gg^{\ggs}_{\gga \ggg}{\cal F}_{\ggb \ggs}-
\delta \Gg^{\ggs}_{\ggb \ggg}{\cal F}_{\ggs \gga}- \nonumber \\
\delta \Gg^{\ggs}_{\gga \ggb}{\cal F}_{\ggg \ggs}- 
\delta \Gg^{\ggs}_{\gga \ggg}{\cal F}_{\ggs \ggb}- 
\delta \Gg^{\ggs}_{\ggb \ggg}{\cal F}_{\gga \ggs}=0 
\end{eqnarray}
Note that the six last terms on the left hand side are second order terms.

	In the absence of electric fields, from the $[0,0,0]$, $[0,0,i]$, 
$[0,i,0]$ and
	$[0,i,i]$ components, we obtain just $0=0$. From $[0,i,j]$
\be
\label{c0ij}
\frac{\pder}{\pder t}{\cal F}_{ij}=0
\ee
and for $[0,1,2]$ (for instance)
\be
\label{c12}
\frac{\pder (R^2 B_3)}{\pder t}=0
\ee

	From $[i,j,k]$
\begin{eqnarray}
\label{cijk}
\frac{\pder}{\pder x^i}{\cal F}_{jk}+
\frac{\pder}{\pder x^j}{\cal F}_{ki}+
\frac{\pder}{\pder x^k}{\cal F}_{ij}- \nonumber \\
\Gg^{0}_{ki}{\cal F}_{j0}-
\Gg^{0}_{ij}{\cal F}_{0k}-
\Gg^{0}_{ji}{\cal F}_{k0}- \nonumber \\
\Gg^{0}_{jk}{\cal F}_{0i}-
\Gg^{0}_{ki}{\cal F}_{0j}-
\Gg^{0}_{kj}{\cal F}_{i0}=0 
\end{eqnarray}
For $[1,2,3]$ for example
\be
\label{c123}
\nabla \cdot \vec{B}=0
\ee
The whole set of equations is reduced to
\be
\label{mfdeq0}
\nabla \cdot \vec{B}=0
\ee
\be
\label{mffc}
\vec{B}R^2={\mbox constant}
\ee
	The first one is familiar. The second one is also familiar but not
obvious. It brings to mind the condition of frozen-in magnetic field lines in 
present cosmic plasmas. What this
equation shows is that the magnetic pattern always remains the same, and
the magnetic strength is reduced by just the effect of expansion. The original
pattern is conserved, even if we are considering small motions of the photon
fluid associated with density inhomogeneities. Neither do small metric fluctuations
alter the magnetic pattern. This property is also obtained for the linear
Newtonian post-Recombination epoch as shown in Appendix ({\ref{app2}).

	From the first set of Maxwell equations
\be
\label{fsome}
{\cal F}^{\gga \ggb}_{~~;\gga}=0
\ee
with the above procedure and the same change of nomenclature, we obtain
\begin{eqnarray}
\label{pfsome}
\frac{\pder}{\pder x^{\gga}}{\cal F}^{\gga \ggb}+
(\delta \Gg^{\gga}_{\ggs \gga}){\cal F}^{\ggs \ggb}+
\Gg^{\gga}_{\ggs \gga}{\cal F}^{\ggs \ggb}+ & &  \nonumber \\
(\delta \Gg^{\ggb}_{\gga \ggs}){\cal F}^{\gga \ggs}+
\Gg^{\ggb}_{\gga \ggs}{\cal F}^{\gga \ggs} & = & 0
\end{eqnarray}
The second and the fourth terms are second order terms, which were included in
the derivation. One of them vanishes, that is $(\delta \Gg^{\ggb}_{\gga \ggs}){\cal F}^{\gga \ggs}=0$.
The result is
\be
\label{pfsome21}
0 = 4\pi R^2 J^0 
\ee
\be
\label{pfsome22}
R^2 4\pi \vec{J}=\nabla \times \vec{B}-\frac{1}{2R^4}
\left[ (\nabla h) \times \vec{B} \right]
\ee

As $[(\nabla h) \times \vec{B}]$
is negligible, we simply obtain that
\be
\label{jo0}
J^0=0
\ee
i.e. macroscopic neutrality, a familiar result, and that
\be
\label{ec} 
\vec{J}=\frac{1}{4\pi R^2} \nabla \times \vec{B}
\ee
which simply provides $\vec{J}$, the electric current, when $\vec{B}$ has
been calculated.

\section{The perturbed equation of motion-energy} 
\label{sec5}
	We start with the equation of motion
\be
\label{eom}
\Tau^{\gga \ggb}_{~~;\ggb}=0
\ee
where
\begin{eqnarray}
\label{tabdef}
\Tau^{\gga \ggb}=pg^{\gga \ggb}+4\pi U^{\gga}U^{\ggb}+ \nonumber \\
\frac{1}{4\pi}
\left(
\begin{array}{cc}
0	&	0	\\
0	&	-R^{-2}\vec{B}\vec{B}
\end{array}
\right)+ 
\frac{1}{4\pi}
\left(
\begin{array}{cc}
\frac{B^2}{2}	&	0 	\\
0		&	\frac{B^2}{2R^2}\delta_{3}
\end{array}
\right)
\end{eqnarray}
The electromagnetic contribution is as in equation (\ref{emmet}) but now taking into
account $\vec{E}=0$.
We now obtain 
\begin{eqnarray}
\label{emcont}
0=\dpar{p}{x^{\ggb}}g^{\gga \ggb}+p\dpar{g^{\gga \ggb}}{x^{\ggb}}+ \nonumber \\
4\dpar{p}{x^{\ggb}}U^{\gga}U^{\ggb}+4p\dpar{U^{\gga}}{x^{\ggb}}U^{\ggb}+ \nonumber \\
4p\dpar{U^{\ggb}}{x^{\ggb}}U^{\gga}+\dpar{\Tau^{\gga \ggb}_{\M}}{x^{\ggb}}+ \nonumber \\
\cris{\gga}{\ggs}{\ggb}\Tau^{\ggs \ggb}_{\M}+
\cris{\ggb}{\ggs}{\ggb}\Tau^{\gga \ggs}_{\M}+ \nonumber \\
p\cris{\gga}{\ggs}{\ggb}g^{\ggs \ggb}+ 
p\cris{\ggb}{\ggs}{\ggb}g^{\gga \ggs}+ \nonumber \\
4p\cris{\gga}{\ggs}{\ggb}U^{\ggs}U^{\ggb}+
4p\cris{\ggb}{\ggs}{\ggb}U^{\gga}U^{\ggs}
\end{eqnarray}
which yields
\begin{eqnarray}
\label{largest}
0=\dpar{\delta p}{x^{\ggb}}g^{\gga \ggb}+
\dpar{p}{x^{\ggb}}h^{\gga \ggb}+ \nonumber \\
(\delta p)\dpar{}{x^{\ggb}}g^{\gga \ggb}+
p\dpar{h^{\gga \ggb}}{x^{\ggb}}+ \nonumber \\
4\dpar{\delta p}{x^{\ggb}}U^{\gga}U^{\ggb}+
4\dpar{p}{x^{\ggb}}(\delta U^{\gga})U^{\ggb}+ \nonumber \\
4\dpar{p}{x^{\ggb}}U^{\gga}\delta U^{\ggb}+
4(\delta p)\dpar{U^{\gga}}{x^{\ggb}}U^{\ggb}+ \nonumber \\
4p\dpar{\delta U^{\gga}}{x^{\ggb}}U^{\ggb}+
4p\dpar{U^{\gga}}{x^{\ggb}}\delta U^{\ggb}+ \nonumber \\
4(\delta p)\dpar{U^{\ggb}}{x^{\ggb}}U^{\gga}+
4p\dpar{\delta U^{\ggb}}{x^{\ggb}}U^{\gga}+ 
4p\dpar{U^{\ggb}}{x^{\ggb}}\delta U^{\gga}+ \nonumber \\
\dpar{\Tau^{\gga \ggb}_{\M}}{x^{\ggb}}+
\cris{\gga}{\ggs}{\ggb}\Tau^{\ggs \ggb}_{\M}+
\cris{\ggb}{\ggs}{\ggb}\Tau^{\gga \ggs}_{\M}+ \nonumber \\
(\delta p)\cris{\gga}{\ggs}{\ggb}g^{\ggs \ggb}+
p(\delta \cris{\gga}{\ggs}{\ggb})g^{\ggs \ggb}+ \nonumber \\ 
p\cris{\gga}{\ggs}{\ggb}h^{\ggs \ggb}+
(\delta p)\cris{\ggb}{\ggs}{\ggb}g^{\gga \ggs}+ \nonumber\\
p(\delta \cris{\ggb}{\ggs}{\ggb})g^{\gga \ggs}+
p\cris{\ggb}{\ggs}{\ggb}h^{\gga \ggs}+ \nonumber \\
4(\delta p)\cris{\gga}{\ggs}{\ggb}U^{\ggs}U^{\ggb}+
4p(\delta \cris{\gga}{\ggs}{\ggb})U^{\ggs}U^{\ggb}+ \nonumber \\
4p\cris{\gga}{\ggs}{\ggb}(\delta U^{\ggs})U^{\ggb}+
4p\cris{\gga}{\ggs}{\ggb}U^{\ggs}\delta U^{\ggb}+ \nonumber \\
4(\delta p)\cris{\ggb}{\ggs}{\ggb} U^{\gga}U^{\ggs}+ 
4p(\delta \cris{\ggb}{\ggs}{\ggb})U^{\gga}U^{\ggs}+ \nonumber \\
4p\cris{\ggb}{\ggs}{\ggb}(\delta U^{\gga})U^{\ggs}+ 
4p\cris{\ggb}{\ggs}{\ggb}U^{\gga}\delta U^{\ggs} 
\end{eqnarray}
which yields for the $0$th component
\begin{eqnarray}
\label{c0}
0=3\dpar{\delta p}{t}+12(\delta p)\frac{\dot{R}}{R}+
4p\nabla \cdot \vec{u}+ \nonumber \\
\frac{2p}{R^2}\dpar{h}{t}- 
4\frac{\dot{R}}{R^3}ph+\dpar{}{t}\frac{B^2}{8\pi}+
4\frac{\dot{R}}{R}\frac{B^2}{8\pi}
\end{eqnarray}
which is the equation of energy conservation. For the $i$th component:
\begin{eqnarray}
\label{ci}
\nabla \delta p+
4R^2\dpar{p}{t}\vec{u}+
4pR^2\dpar{\vec{u}}{t}- 
\frac{1}{4\pi}\vec{B}\cdot \nabla \vec{B}+ \nonumber \\
\nabla \frac{B^2}{8\pi}+
p \left( \frac{1}{R^2}+R^2 \right)\nabla \cdot \vec{\vec{h}}+
20pR\dot{R}\vec{u}=0
\end{eqnarray}

	However, it is possible to show that $\nabla \cdot \vec{\vec{h}}=0$. This
comes from the well known property of the tensor metric $g_{\gga \ggb; \ggg}=0$,
and therefore $g_{\gga \ggb; \ggb}=0$ and $g_{i \ggb; \ggb}=0$, which becomes
in our case $g_{i \ggb; \ggb}+h_{i \ggb; \ggb}$ where $g_{i \ggb}$ now corresponds
to the Robertson-Walker metric. We have
\be
\label{gibb1}
g_{i \ggb; \ggb}=\dpar{g_{i \ggb}}{x^{\ggb}}-
\cris{\ggs}{\ggb}{i}g_{\ggs \ggb}-
\cris{\ggs}{\ggb}{\ggb}g_{i \ggs}=0
\ee
hence
\be
\label{hibbeq0}
h_{i \ggb; \ggb}=0
\ee
and
\be
\label{phiixjeq0}
0=\dpar{h_{i \ggb}}{x^{\ggb}}-
\cris{\ggs}{\ggb}{i}h_{\ggs \ggb}-
\cris{\ggs}{\ggb}{\ggb}h_{i \ggs}=
\dpar{h_{ij}}{x^j}
\ee
With this simplification and some algebra, we have for the $i$th component
\be
\label{ci2}
\dpar{}{t}(4pR^5\vec{u})+R^3 \left[ 
\nabla \left( \delta p+\frac{B^2}{8\pi} \right)
-\frac{1}{4\pi} \vec{B} \cdot \nabla \vec{B} \right]=0
\ee
which is the equation of motion.

\section{The perturbed Einstein field equations}
\label{sec6}
	We start with the Einstein field equations in the form
\be
\label{efeq1}
R_{\ggm \ggn}=-8\pi(\Tau_{\ggm \ggn}-\frac{1}{2}g_{\ggm \ggn}\Tau^{\ggl}_{~\ggl})
\ee
where $\Tau^{\gga \ggb}$ is given by equation (\ref{tabdef}). For the electromagnetic
contribution we have 
\be
\label{emconttot}
4\pi\Tau_{\M \gga \ggb}=
\left(
\begin{array}{cc}
0       &       0       \\
0       &       -R^{2}\vec{B}\vec{B}
\end{array}
\right)+
\left(
\begin{array}{cc}
\frac{B^2}{2}   &       0       \\
0               &      R^2 \frac{B^2}{2}\delta_{3}
\end{array}
\right)
\ee
and
\be
\label{tmlleq0}
\Tau^{~\ggl}_{\M ~\ggl}=0
\ee
$R_{\ggm \ggn}$ is Ricci's tensor, where
\begin{eqnarray}
\label{rten}
\delta R_{ij} & = & \frac{1}{2R^2}\left[ 
\nabla ^2 \vec{\vec{h}}+\nabla \nabla \vec{\vec{h}}
\right]-
\frac{1}{2}\dpar{^2\vec{\vec{h}}}{t^2}+ \nonumber \\
 & & \frac{\dot{R}}{2R} \left[
\dpar{\vec{\vec{h}}}{t}-\dpar{h}{t}\vec{\vec{\delta}} \right]+
\frac{{\dot{R}}^2}{R^2}[-2\vec{\vec{h}}+h\vec{\vec{\delta}}] \\
\delta R_{0i} & = & \frac{1}{2}\dpar{}{t}[R^{-2}\nabla h] \\
\delta R_{00} & = & \frac{1}{2R^2}\left[
\dpar{^2 h}{t^2}-2\frac{\dot{R}}{R}\dpar{h}{t}+2 \left(
\frac{{\dot{R}}^2}{R^2}-\frac{\ddot{R}}{R} \right) h \right]
\end{eqnarray}
as in Weinberg (1972). Then, we obtain
\begin{eqnarray}
\label{pfe1}
\nabla ^2 \vec{\vec{h}}+\nabla \nabla \vec{\vec{h}}-
R^2\dpar{^2\vec{\vec{h}}}{t^2}+ & & \nonumber \\
R\dot{R}\left( \dpar{\vec{\vec{h}}}{t}- 
\dpar{h}{t}\vec{\vec{\delta}} \right)
+2{\dot{R}}^2 (-2\vec{\vec{h}}+h\vec{\vec{\delta}})+ & & \nonumber \\
+8\pi pR^2 \vec{\vec{h}}+ 8\pi (\delta p) R^2 \vec{\vec{\delta}}- & & \nonumber \\
4R^4\vec{B}\vec{B}+2R^4B^2\vec{\vec{\delta}} & = & 0
\end{eqnarray}
\be
\label{pfe2}
\dpar{}{t}\left( \frac{\nabla h}{R^2} \right) = 64\pi pR^2\vec{u}
\ee
\begin{eqnarray}
\label{pfe3}
\dpar{^2 h}{t^2}-2\frac{\dot{R}}{R}\dpar{h}{t}+2 \left(
\frac{{\dot{R}}^2}{R^2}-\frac{\ddot{R}}{R} \right) h+ & & \nonumber \\
48\pi(\delta p)R^2+2B^2R^2 & = & 0
\end{eqnarray}

\section{Density perturbations}
\label{sec7}
	Let us define the relative perturbed density as:
\be
\label{deltadef}
\delta=\frac{\delta p}{p}=\frac{\delta \gge}{\gge}
\ee
($\delta$ is not so defined by other authors. Some
others adopt $\delta=\delta \gge /(\gge+p)$ which becomes in this case $3/4$ times
the value of our $\delta$). The final purpose is to deduce $\delta(t, \vec{r})$, i.e. how
density inhomogeneities evolve, under the action of gravity and an unperturbed
underlying magnetic configuration. To find the differential equation for $\delta$,
we must perform some algebra because our set of equations is still rather large
and complicated. We must first choose which equations will be handled, because 
as stated before, they are not all independent. Our unknowns are $\delta p$, 
$\vec{u}$, $\vec{\vec{h}}$, as $\vec{B}$ has already been determined as a
function of time. We therefore have ten unknowns. The Einstein field equations are 
also made up of ten equations, and from them the equations of motion and energy are derivable.
On the other hand, Maxwell's equations have already been used to deduce the 
evolution of the magnetic field and the electrical current. Though we could
therefore use only the Einstein field equations, the final objective is straightforwardly
reached with the equation of motion (\ref{ci2}), the equation of energy balance (\ref{c0}),
and equations (\ref{pfe2}) and (\ref{pfe3}) from the whole set of Einstein field equations, selected
because they do not contain the tensor $\vec{\vec{h}}$ but just its trace $h$.
The unknowns are now $\delta p$, $\vec{u}$ and $h$. We still have more 
equations than unknowns. Let us rewrite the equations with the introduction of
``present day'' quantities, with the subindex $0$.
\be
\label{pp0}
p_0=pR^4
\ee
\be
\label{bb0}
\vec{B_0}=\vec{B}R^2
\ee
\be
\label{dpdp0}
\delta p_0=(\delta p)R^4
\ee
\be
\label{hh0}
h_0=hR^{-2}
\ee

	Here $p_0$ and $\vec{B_0}$ would actually be the present values of $p$
and $\vec{B}$ if our equations were valid for all subsequent periods.  
The law $pR^4={\mbox {\em constant}}$ is valid for photons in the Acoustic and Post-Recombination
eras. However $\vec{B_0}$ has recently been affected by non-linear effects and
probably does not coincide with the present magnetic field. While $p_0$ and
$\vec{B_0}$ are constant, $\delta p_0$ and $h_0$ are not. Neither $\delta p_0$
nor $h_0$ represent actual present values, but they are preferred as variables because in the
variations of $\delta p_0$ and $h_0$ the effect of pure expansion is suppressed.
We then have:
\be
\label{mov}
4p_0R\dpar{}{t}(R\vec{u})+\nabla \delta p_0-
\frac{1}{4\pi}\vec{B_0} \cdot \nabla \vec{B_0}+
\nabla \frac{B^{2}_{0}}{8\pi}=0
\ee
\be
\label{cont}
3\dpar{\delta p_0}{t}+4p_0\nabla \cdot \vec{u}+
2p_0\dpar{h_0}{t}+\dpar{}{t} \left( \frac{B^{2}_{0}}{8\pi} \right)=0
\ee
\be
\label{met1}
R^2\dpar{}{t} \nabla h_0=64\pi p_0 \vec{u}
\ee
\be
\label{met2}
R^2\dpar{}{t}\left( R^2 \dpar{h_0}{t} \right) + 48\pi \delta p_0+
16\pi \frac{B^{2}_{0}}{8\pi}=0
\ee
As suggested by the equations themselves, we now change the temporal and
spatial variables, and use the definition (\ref{deltadef}) of $\delta$. As
new spatial variables we will use
\be
\label{svar}
x'_i=\frac{K}{R_e}x_i=2.73\times10^{-15}x_i
\ee
where $K$ is the constant in the expansion law
\be
\label{explaw}
R^2=Kt
\ee
in the radiation-dominated era. Its value is
\be
\label{explawconst}
K=4\sqrt{2\pi p_0}=2.73\times10^{-20} s^{-1}
\ee
and $R_e$ is the cosmological scale factor at the last time in the period
considered in this paper, which is close to Equality. For $p_0$ we have
adopted the value $8.84\times 10^{-42} s^{-2}$. It should be noted 
that $x_i$ are comoving coordinates measured in seconds. They now coincide
with present coordinates. (It is easily obtained that $1~Mpc=1.03\times 10^{14} s$).
However $x'_i$ is dimensionless
\be
\label{xpd}
x'_i=0.28 \left( \frac{d}{Mpc} \right)
\ee
where $d$ is the distance in $Mpc$.

	As a temporal variable let us choose
\be
\label{tvar}
\ggt=\ln \frac{t_e}{t}=2\ln \frac{R_e}{R}
\ee
and therefore
\be
\label{dtdtaurel}
-\frac{R^2}{dt}=\frac{K}{d\ggt}
\ee
where $t_e$ is the last time of the period considered here, close to Equality,
corresponding to $R=10^{-5}$, i.e. $3.66\times 10^{9} s$. $\ggt$ is a time
variable increasing backwards from future to past. We then introduce some
time independent functions defining the magnetic pattern:
\be
\label{Xdef}
X=\frac{B^{2}_{0}}{24\pi p_0}
\ee
\be
\label{ndef}
\vec{n}=-\frac{\vec{B_0} \cdot \nabla ^\prime \vec{B_0}}{4\pi p_0}+3\nabla ^\prime X
\ee
\be
\label{mdef}
2m=-\nabla ^\prime \cdot \left( \frac{2}{3}\vec{n}-\nabla ^\prime X \right)
\ee
where $\nabla ^\prime$ is the gradient using $x'_i$ instead of $x_i$. In the
unit system we are using $1~{\mbox {\em Gauss}}=8.61\times 10^{-15} s^{-1}$.
In this way all the magnitudes of the different quantities involved are close
to unity and the equations become extremely simple.

	We also change the nomenclature, so that a point over a quantity's symbol
now means its derivative with respect to $\ggt$, instead of with respect to $t$.
Thus we have
\be
\label{newmov}
2R_e\vec{u}-4R_e\dot{\vec{u}}+\nabla ^\prime \delta+\vec{n}=0
\ee
\be
\label{newcont}
-3\dot{\delta}+4R_e e^{-\ggt} \nabla ^\prime \cdot \vec{u}-2\dot{h_0}=0
\ee
\be
\label{newmet1}
2\ddot{h_0}+3\delta+3X=0
\ee
\be
\label{newmet2}
\nabla ^\prime \dot{h_0}+2R_e\vec{u}=0
\ee

	Taking $\nabla ^\prime$ in equation (\ref{newmet1}) and $\dpar{}{\ggt}$ in equation (\ref{newmet2})
and subtracting, inserting this $\dot{\vec{u}}$ in equation (\ref{newmov}) to obtain
$\vec{u}$, calculating $\dpar{}{\ggt}$ in equation (\ref{newcont}), taking into
account the obtained functions $\vec{u}$ and $\dot{\vec{u}}$ 
in equation (\ref{newmet1}), we obtain a differential equation containing only $\delta$ as
a variable, $\delta(\ggt, x' _i)$
\be
\label{deltaeq}
\ddot{\delta}-\delta-X+\frac{1}{3}e^{-\ggt}\nabla ^{\prime 2} \delta+
2e^{-\ggt}m=0
\ee
This is our basic equation, an elliptic linear second order differential equation
with variable coefficients. The complete discussion may have complications arising
from its elliptic nature (if we prefer not to choose a boundary condition for $\ggt=0$).
The following section contains a preliminary insight.

\section{Conclusions}
\label{sec9}

	As previously mentioned, the implications of various individual
magnetic patterns and of a cosmological magnetic spatial spectrum in the
formation of structures will be studied in forthcoming papers. In particular,
we will examine in paper II the effects of a magnetic flux tube, the simplest
structure with cylindrical symmetry. Our basic results will be confirmed with
this simple particular case. We will now examine the general results obtainable
from the differential equation (\ref{deltaeq}).

	The primordial magnetic field spatial spectrum remains time independent
in the linear approximation considered here,
being diluted just by the general expansion but conserving patterns and relative
values. 
 Therefore, present large scale magnetic field patterns in the 
intergalactic medium, if we are eventually able to observe them, could reveal
very early processes concerning magnetogenesis. Complementary to the standard
methods reviewed by Kronberg (1994), the measurement of the highest energy
cosmic rays (Lee, Olinto, \& Siegl, 1996) and other indirect methods 
(Plaga 1995; 
 Kronberg 1995; Battaner et al. 1991) are promising in the future. 
There is even the possibility of measuring magnetic fields at the last
scattering surface (Kosowsky \& Loeb 1996).

	The fact that magnetic inhomogeneities are conserved until recent epochs
in the history of the Universe supports the works by Wasserman(1978) and
by Kim et al. (1994) in which a preexisting magnetic configuration
determines the evolution of density inhomogeneities.
The influence of this primordial magnetic field spectrum on the formation
of large-scale structures, clusters and superclusters, as well as of galaxies, is
very important.

	If magnetic fields have enough strength they can have a clear influence
on the origin and evolution of density inhomogeneities. For very early times,
with very large $\tau$, or equivalently for very large scale structures, the
last two terms in eq. (\ref{deltaeq}) would become smaller and the resulting equation would be easily
integrated to give

\be
\label{new}
\delta=-X+c_1 e^{-\tau}+c_2 e^{\tau}
\ee

	This equation will be analyzed in detail in paper II. We now see that
for very large $\tau$, $\delta$ would become very large unless $c_2 =0$. For
large $\tau$ we would then have $\delta=-X$.

	Observe that when $X=0$ this equation is just the solution in the
absence of magnetic fields, expressing the growth of the compressional mode as
summarized in the classic book by Weinberg (1972), giving $\delta \propto t
\propto R^2$.

Then, magnetic fields may provide an alternative mechanism to generate
the promordial spectrum of density inhomogeneities, which are then amplified by 
gravitational instabilities. Other primordial mechanisms, such as quantum
fluctuations at Inflation, or exotic discontinuities produced by phase
transitions, cannot be rejected, but magnetism is a very interesting additional
possibility
as, if $\delta$ were null initially, and only the homogeneous Universe were
perturbed by random magnetic fields, $\delta$ would have been created.

	On the other hand, not only are primordial magnetic fields able to
originate density structures at very early epochs, but they have a direct influence
on the evolution of structures at $R \approx 10^{-5}R_0$. The effect of this
direct influence may be of a very different nature, depending on the initial
conditions and on the type of the magnetic field pattern, producing
concentrations of photons which would eventually be the site
of baryon concentrations.

	The term $X$ only depends on the magnetic field energy density, but the
last term, including $m$, depends on the magnetic field as a vector, thus
producing anisotropies in the evolution of density inhomogeneities. This
anisotropic evolution would be more important at relative recent times, close
to Equality, and for small scale structures. Later on, heat conduction and
viscosity would eliminate small scale structures and, after Recombination,
non-linear effects would complicate the initial structure, but
the primordial distribution of magnetic fields and that of the density at the
epoch of Equality would probably still be recognizable at present.

	It is important to obtain the order of magnitude of the field strength
able to affect the evolution of the density inhomogeneities. The magnitude of
$X$ must be at least of the same order than $\delta$, which can have typical
values of $10^{-5}$ in the period of time considered. If $X$ were unity, we
would have equipartition between magnetic and radiative energy densities and we
would have $B_0 \approx 3 \ggm G$. But if $X \approx 10^{-5}$, we see from its
definition ($X$ depends on $B_0^2$) that $B_0 \approx 10^{-8} G$. Therefore,
magnetism affects the evolution of the density pattern if $B_0 \gsim 10^{-8}
G$. Higher fields are not expected as they would produce a very fast density
evolution, incompatible with present observations. The magnetic field may not
differ very much from $B_0 \approx 10^{-8} G$.

	This is an equivalent-to-present value. Real magnetic fields at any
time are calculated taking into account $BR^2=B_0^2$. It is important to
calculate $B$ at the epoch of nucleosynthesis. With $R=2.3 \times 10^{-9}$ at
the end of nucleosynthesis, we have $B_N \approx 2 \times 10^9 G$, perfectly
compatible with the limits of Grasso and Rubinstein (1995) of $B_N \leq 3
\times 10^{10} G$.

\appendix

\section{Mean magnetic field in the Robertson-Walker metric}
\label{app1}
	The purpose here is to show that the mean magnetic field is null in a
universe obeying the metric of Robertson-Walker, in order to justify this
assumption in the paper.

	First consider the Maxwell equations:
\be
\label{ap1em1}
{\cal F}^{\gga \ggb}_{~~; \gga }=-4\pi J^{\ggb}
\ee
\be
\label{ap1em2}
{\cal F}_{\ggb \ggg ; \gga}+
{\cal F}_{\ggg \gga ; \ggb}+
{\cal F}_{\gga \ggb ; \ggg}=0
\ee
where now (in contrast with the nomenclature above used) 
${\cal F}^{\gga \ggb}$ and $J^{\ggb}$ are the unperturbed Faraday
tensor (a function of the unperturbed mean electric and magnetic fields
$\vec{E}$ and $\vec{B}$) and the unperturbed charge density electric current
vector. All these quantities have been assumed to be null throughout the paper.
We would now have, from (\ref{ap1em2})
\begin{eqnarray}
\label{ap1long}
0=\dpar{{\cal F}_{\ggb \ggg}}{x^{\gga}}+
\dpar{{\cal F}_{\ggg \gga}}{x^{\ggb}}+
\dpar{{\cal F}_{\gga \ggb}}{x^{\ggg}}- \nonumber \\
\cris{\ggs}{\gga}{\ggb} {\cal F}_{\ggs \ggg}-
\cris{\ggs}{\gga}{\ggg} {\cal F}_{\ggb \ggs}-
\cris{\ggs}{\ggb}{\ggg} {\cal F}_{\ggs \gga}- \nonumber \\
\cris{\ggs}{\gga}{\ggb} {\cal F}_{\ggg \ggs}-
\cris{\ggs}{\gga}{\ggg} {\cal F}_{\ggs \ggb}-
\cris{\ggs}{\ggb}{\ggg} {\cal F}_{\gga \ggs} 
\end{eqnarray}
From this equation we obtain
\be
\label{evar}
\dpar{\vec{E}}{t}(1-R^2)=2 \frac{\dot{R}}{R} \vec{E}
\ee
\be
\label{dive0}
\nabla \cdot \vec{E}=0
\ee
\be
\label{bvar}
\dpar{(R^2 B_3)}{t}+\dpar{(R^2 E_2)}{x^1}-\dpar{E_1}{x^2}=0
\ee
(and other similar formulae)
\be
\label{eeq0}
\vec{E}=0
\ee
\be
\label{divb0}
\nabla \cdot \vec{B}=0
\ee
	With equation (\ref{eeq0}) in (\ref{evar}), (\ref{dive0}) and (\ref{bvar}) we have
\be
\label{fluxcons}
\vec{B} R^2={\mbox {\em constant}}
\ee
Therefore the Maxwell equations under a Robertson-Walker metric are not incompatible
with a mean magnetic field. The other set of Maxwell equations simply tells us that
the Universe should be macroscopically neutral and that electric currents 
given by
\be
\vec{J}=\frac{1}{4\pi R^2} \nabla \times \vec{B}
\label{ecurr1}
\ee
exist. Neither is the law of conservation of momentum-energy very restrictive
with respect to the existence of a cosmological magnetic field. We can similarly 
benefit from the derivation carried out in section (\ref{sec5}). Now, again, $\vec{B}$ is 
the unperturbed magnetic field. For energy conservation we would have
\be
\dpar{\gge}{t}+\dpar{}{t} \left( \frac{B^2}{8\pi} \right) + 
4\frac{\dot{R}}{R} \gge + 4 \frac{\dot{R}}{R}\frac{B^2}{8\pi}=0 
\label{energycons}
\ee
where $\gge$  is the photon energy density. If $x^1$ is the direction of the
magnetic field, we have for the conservation of momentum, taking into
account the homogeneity of the Universe $\dpar{p}{x^i}=0$
\be
\label{hom1}
\nabla \frac{B^2}{2} = \nabla \cdot (\vec{B} \vec{B})
\ee
which does not imply $\vec{B}=0$ at all. This result is however, deduced from the 
Einstein field equations. We again benefit from the above calculations, with $\vec{B}$
now being the unperturbed magnetic field. It is obtained that
\begin{eqnarray}
\label{fourmat}
\left(
\begin{array}{cc}
3\frac{\ddot{R}}{R}	&	0	\\
0			&	-(R\ddot{R}+2\dot{R}^2)\delta _3
\end{array}
\right) =-2\left(
\begin{array}{cc}
0	&	0	\\
0	&	-R^2\vec{B}\vec{B}
\end{array}
\right) \nonumber \\
- \left(
\begin{array}{cc}
B^2	&	0	\\
0	&	R^2 B^2 \delta _3
\end{array}
\right) - 8\pi \left(
\begin{array}{cc}
\gge	&	0	\\
0	&	pR^2 \delta _3
\end{array}
\right) 
\end{eqnarray}
Components $(1, 2)$, $(1, 3)$ and $(2, 3)$ of this equation yields
\be
\label{b12}
B_1 B_2 =0
\ee
\be
\label{b13}
B_1 B_3 =0
\ee
\be
\label{b23}
B_2 B_3 =0
\ee
This does not mean that $B_1 =B_2 =B_3 =0$. One of them may be non-vanishing,
for instance $B_1$. But then components $(1, 1)$ and $(2, 2)$ yield
\be
\label{comp11}
-(R\ddot{R}+2\dot{R}^2)=2R^2 B^2_1 - R^2 B^2_1 - 8\pi p R^2
\ee
\be
\label{comp22}
-(R\ddot{R}+2\dot{R}^2)=- R^2 B^2_1 - 8\pi p R^2
\ee
and subtracting we obtain $B_1 =0$, and therefore
\be
\label{bis0}
\vec{B}=0
\ee
in a Robertson-Walker metric.

	We conclude that in a universe with a Robertson-Walker metric, the
existence of $\vec{B}$, even if compatible with the Maxwell equations and with
the equations of conservation of energy-momentum, is incompatible with the
Einstein field equations.

	From (\ref{fourmat}) it is then obtained
\be
\label{newa1}
\frac{\ddot{R}}{R}+8\pi p+ B^2=0
\ee
\be
\label{newa2}
R\ddot{R}+2\dot{R}^2=8 \pi p R^2 + \frac{B^2}{3} R^2
\ee
Eliminating $\ddot{R}$ in these two equations it is obtained
\be
\label{newa3}
\frac{8 \pi \gge R^3}{3}+\frac{1}{3}B^2 R^2=\dot{R}^2 R
\ee
which is the generalization of the Einstein-Friedmann equation, and which was
implicitly used when obtaining (\ref{r2per4ht}).

\section{Linear Newtonian evolution of magnetic fields}
\label{app2}
	It will be shown that in the linear Post-Recombination epoch, the
magnetic structures evolve according to the law
$\vec{{\cal B}}R^2={\mbox {\em constant}}$, with $\vec{{\cal B}}$ being the amplitude of the
magnetic propagating inhomogeneity (we are not now using comoving coordinates)
in a similar way to their evolution in the Radiation dominated epoch.

Let us consider the induction equation
\be
\label{induction}
\dpar{\vec{B}}{t}=\nabla \times (\vec{v} \times \vec{B} )=
\vec{B} \cdot \nabla \vec{v} - \vec{v} \cdot \nabla \vec{B}-
\vec{B} \nabla \cdot \vec{v}
\ee
As the mean field vanishes, let us again term
$\delta \vec{B} \rightarrow \vec{B}$. After the introduction of perturbations,
this becomes
\be
\label{pertinduc}
\dpar{\vec{B}}{t}=\vec{B} \cdot \nabla \vec{v} - \vec{v} \cdot \nabla \vec{B}-
\vec{B} \nabla \cdot \vec{v}
\ee
where $\vec{v}$ is the mean velocity. We would have in general 
$\vec{v} \rightarrow \vec{v} + \vec{u}$, with $\vec{u}$ being the fluctuative
velocity, but it would be present only in second order terms. Therefore 
$\vec{v}$ obeys
\be
\label{vvsr}
\vec{v}=\frac{\dot{R}}{R}\vec{r}
\ee
where $\vec{r}$ is the position vector. With
\be
\label{divv}
\nabla \cdot \vec{v} = 3\frac{\dot{R}}{R}
\ee
\be
\label{gradv}
\nabla \vec{v}=\frac{\dot{R}}{R} \vec{\vec{\delta}}
\ee
we obtain
\be
\label{newind}
\dpar{\vec{B}}{t}=-2\frac{\dot{R}}{R}\vec{B}-
\frac{\dot{R}}{R} \vec{r} \cdot \nabla \vec{B}
\ee
The solution would be of the form
\be
\label{bb}
\vec{B}=\vec{{\cal B}}{\cal E}
\ee
where $\vec{{\cal B}}$ is the amplitude that only depends on time, and
${\cal E}$ is a function only of $\vec{r}$:
\be
\label{efour}
{\cal E}=e^{i\vec{q} \cdot \frac{\vec{r}}{R}}
\ee
where $\vec{q}$ is constant. Substituting in the induction equation, we have
\be
\label{last3}
\dot{\vec{{\cal B}}}+
2\frac{\dot{R}}{R}\vec{{\cal B}}-
i\frac{\vec{q} \cdot \vec{r}}{R^2}\dot{R} \vec{{\cal B}}+
i\frac{\vec{q} \cdot \vec{r}}{R^2}\dot{R} \vec{{\cal B}}=0
\ee
hence
\be
\label{last2}
\dot{\vec{{\cal B}}}+
2\frac{\dot{R}}{R}\vec{{\cal B}}=0
\ee
and
\be
\label{last}
\vec{{\cal B}} R^2={\mbox {\em constant}}
\ee

	It can therefore be concluded that the magnetic spatial primordial
pattern has been preserved until very recently when the back-reaction of the
velocity field onto the magnetic field became important. The importance
of the back-reaction has been pointed out by Kim et al. (1994).


\begin{thebibliography}{}
\bibitem{Batt1996} Battaner E. 1996, {\em Astrophysical Fluid Dynamics}, Cambridge Univ. Press. Cambridge
\bibitem{Batt1991} Battaner E., Garrido J.L., S\'anchez-Saavedra M. L., \& Florido E. 1991,
\aap , 251, 402
\bibitem{Born1988} B\"{o}rner G. 1988, {\em The early Universe}, Springer-Verlag, Berlin
\bibitem{Bran1996} Brandenburg A., Enqvist K., \& Olesen P. 1996, Preprint astro-ph/9602031
\bibitem{Chen1994} Cheng B. \& Olinto A. V. 1994, Phys. Rev. D, 50, 2421
\bibitem{Col1992} Coles P. 1992, Comments Astrophys., 16, 45-60
\bibitem{Dav1996} Davis A. C. \& Dimopoulos K. 1996, Preprint CERN-TH/95-175
\bibitem{Dett1993} Dettmann C. P., Frankel N. E., \& Kowalenko V. 1993, \prd ,  48, 5655
\bibitem{Enqv1993} Enqvist K. \&  Olesen P. 1993, Phys. Lett. B, 319, 178
\bibitem{Enqv1994} Enqvist K. \& Olesen P. 1994, Nordita preprint 94/6 P 
\bibitem{Fer1995} Feretti L., Dallacasa, Giovannini G., \& Tagliani A. 1995, 
  \aap , 302, 680
\bibitem{Fiel1969} Field G. B. 1969, in {\em Astrophysics and General Relativity}, Vol I,
ed. by M. Chretien, S. Deser \& J. Goldstein. Gordon and Brench. New York, pg 59
\bibitem{Gail1994} Gailis R. M., Dettmann C. P., Frankel N. E., \& Kowalenko V. 1994, \prd , 50, 3847
\bibitem{Gail1995} Gailis R. M., Frankel N. E., \& Dettmann C. P. 1995, \prd ,  52, 6901
\bibitem{Gras1995} Grasso D. \& Rubinstein H. R. 1995, Astropart. Phys.,  3, 95
\bibitem{Holc1990} Holcomb K. A. 1990, \apj , 362, 381
\bibitem{Holc1989} Holcomb K. A. \& Tajima T. 1989, \prd , 40, 3809
\bibitem{Kern1995} Kernan, P. J., Starkman G. D., \& Vachaspati T. 1995, Preprint CWRU-P10-95
\bibitem{Kim1994} Kim E., Olinto A., \& Rosner R. 1996, \apj, 468, 28
\bibitem{Kolb1990} Kolb E. W. \& Turner M. S. 1990, {\em The early Universe}, 
Addison-Wesley Pub. Co.
\bibitem{Koso1996} Kosowsky A. \& Loeb A. 1996, \apj, 469, 1
\bibitem{Kron1994} Kronberg P. 1994, {\em Extragalactic magnetic fields}, 
Rep. Progress in Physics  57, 325
\bibitem{Kron1995} Kronberg P.P. 1995, Nature, 374, 404
\bibitem{Lee1996} Lee S., Olinto A. V., \& Sigl G. 1996,  \apjl, 455, 21
\bibitem{Les1995} Lesch H. \& Chiba M. 1995, \mnras , 241, 1
\bibitem{Peeb1980} Peebles P. J. E. 1980, {\em The large-scale structure of the
Universe}, Princeton Series in Physics
\bibitem{Per1988} Peratt A. L. 1988, Laser and Particle Beams, 6, 471
\bibitem{Plag1995} Plaga R. 1995, Nature, 374, 430
\bibitem{Quas1989} Quashnock J. M., Loeb A., \& Spergel D. N. 1989, \apj , 344, L49
\bibitem{Rat1992} Ratra B. 1992, \apj , 391, L1
\bibitem{Rees1987} Rees M. J. 1987, \qjras , 28, 197
\bibitem{Ruz1989} Ruzmaikin A., Sokoloff D., \& Shukurov A. 1989, \mnras , 241, 1
\bibitem{Turn1988} Turner M. S., \& Widrow L. M. 1988, Phys. Rev. D, 37, 2743
\bibitem{Vach1992} Vachaspati T. 1991, Phys. Lett., 265, 258
\bibitem{Wass1978} Wasserman I. 1978, \apj , 224, 337
\bibitem{Wein1972} Weinberg S. 1972, {\em Gravitation and Cosmology}, J. Wiley \& Sons. New York
\bibitem{Welt1984} Welter G. L., Perry J. J., \& Kronberg P. P. 1984, \apj , 279, 19
\bibitem{Wolf1992} Wolfe A. M., Lanzetta K. M., \& Oren A. L. 1992, \apj , 388, 17
\end{thebibliography}
\end{document}